# The Sensitivity of Nucleosynthesis in Type I X-ray Bursts to Thermonuclear Reaction-Rate Variations

Anuj Parikh[a,*,1], Jordi José[a,b], Fermín Moreno[a], Christian Iliadis[c,d]

[a]*Departament de Física i Enginyeria Nuclear, EUETIB, Universitat Politècnica de Catalunya, C./ Comte d'Urgell 187, E-08036 Barcelona, Spain*

[b]*Institut d'Estudis Espacials de Catalunya (IEEC), Ed. Nexus-201, C./ Gran Capità 2-4, E-08034 Barcelona, Spain*

[c]*Department of Physics and Astronomy, University of North Carolina, Chapel Hill, NC 27599-3255, USA*

[d]*Triangle Universities Nuclear Laboratory, Durham, NC 27708-0308, USA*

**Abstract.** We examine the sensitivity of nucleosynthesis in Type I X-ray bursts to variations in nuclear rates. As a large number of nuclear processes are involved in these phenomena -with the vast majority of reaction rates only determined theoretically due to the lack of any experimental information- our results can provide a means for determining which rates play significant roles in the thermonuclear runaway. These results may then motivate new experiments. For our studies, we have performed a comprehensive series of one-zone post-processing calculations in conjunction with various representative X-ray burst thermodynamic histories. We present those reactions whose rate variations have the largest effects on yields in our studies.

## Introduction

Type I X-ray bursts (XRBs) result from thermonuclear runaways in hydrogen- and/or helium-rich material accreted on to the surface of a neutron star in a low-mass X-ray binary system (see recent reviews by Strohmayer and Bildsten (2006); Schatz and Rehm (2006)). Since the discovery of these phenomena by Grindlay et al. (1976) and Belian et al. (1976), over 80 such bursting systems have been found in our Galaxy by X-ray observatories such as the Rossi X-ray Timing Explorer, BeppoSAX, XMM-Newton and

[*]Corresponding Author. Tel. +49 (0)89 289 12489; Fax. +49 (0)89 289 12435; e-mail: anuj.parikh@ph.tum.de

[1]Present Address: Physik Department E12, Technische Universität München, James-Franck-Strasse, D-85748 Garching, Germany

Chandra (Strohmayer and Bildsten, 2006; Galloway et al., 2006); detections have also been made outside the Milky Way (e.g. Pietsch and Haberl (2004)).

XRB simulations have shown that burst properties such as energy ($10^{39}$ – $10^{40}$ ergs, $L_{peak}$~ 3 x $10^{38}$ erg/s), recurrence time (hours to days) and duration (~ 10 -100 s) depend strongly upon characteristics such as accretion rate (~ $10^{-8}$ – $10^{-10}$ $M_{sol}$/yr) and composition of the accreted matter (see e.g. Fujimoto et al. (1981)), as well as the ashes of burning from previous bursts (so-called compositional inertia, see Taam (1980) and Woosley et al. (2004)). Moreover, nucleosynthesis in XRBs will likely also affect the evolution of thermal and electrical properties of the neutron star (e.g. Schatz et al. (1999)), the possibility of a subsequent superburst (unstable ignition of a carbon layer accumulated through several bursts – see e.g. Cumming and Bildsten (2001)), and, perhaps, Galactic abundances. Although ejection of material during an XRB is unlikely due to the deep gravitational potential of a neutron star, it has been shown that radiation-driven winds involved in XRBs exhibiting photospheric-radius-expansion may indeed lead to such ejection (Weinberg et al., 2006). Given an ejection process, nucleosynthesis in XRBs may then have important consequences for the origin of the puzzling light p-nuclei (Schatz et al., 1998).

Because of the extreme conditions during an XRB ($T_{peak} \gtrsim 10^9$ K, $\rho \sim 10^6$ g/cm$^3$), nuclear reactions are driven towards the proton drip-line and to high masses (as far as $^{126}$Xe in Koike et al. (2004), but see also Schatz et al. (2001) and Woosley et al. (2004)) through the $\alpha$p and rp processes (Wallace and Woosley, 1981; Schatz et al., 1998). Computational limitations have forced most previous theoretical studies of XRBs to focus either upon the stellar model by using hydrodynamic simulations (at the expense of detailed reaction networks – e.g. Woosley and Weaver (1984); Taam et al. (1993); Taam, Woosley and Lamb (1996)) or, upon the nuclear physics (through one-zone calculations that neglect, for example, convection – see e.g. Schatz et al. (1999, 2001); Koike et al. (1999)). Recently, however, significant progress has been made towards more realistic models by coupling (1-D) hydrodynamic calculations with large reaction networks: see Woosley et al. (2004) (using an adaptive network with up to 1300 isotopes, ending at Po); Fisker et al. (2004, 2006, 2007, 2008) (using ~ 300 isotopes up to Te); and José and Moreno (2006) (using 478 isotopes up to Te).

**Sensitivity Studies**

Due to the large number of reaction rates involved in XRB nucleosynthesis (most of which can only be estimated theoretically at the moment), it is important to identify which of these rates significantly affect XRB properties such as nucleosynthesis and light curves. Reducing the uncertainties of such rates can help to expose weaknesses in the *astrophysical* assumptions of different models, ideally leading to convergence of XRB model predictions. Characteristics of the observed XRB light curves then provide direct constraints on hydrodynamic models. As well, the ashes of XRBs may show gravitationally-redshifted atomic absorption lines from the neutron star surface, observable through high-resolution X-ray spectra (Cottam et al., 2002; Bildsten et al., 2003; Chang et al., 2005; Weinberg et al., 2006). Such features would directly probe XRB nucleosynthesis.

Most previous attempts to examine the effects of rate uncertainties in XRBs either varied a few specific reactions individually (Iliadis et al., 1999; Thielemann et al., 2001; Fisker et al., 2004, 2006, 2008), or varied groups or entire libraries of rates (Wallace and Woosley, 1981; Schatz et al., 1998; Koike et al., 1999; Woosley et al., 2004). Clearly a more systematic and extensive exploration of the impact of rate uncertainties in XRB models is desirable to guide relevant future experiments.

A natural way to proceed is through sensitivity studies. In general, two techniques can be used: varying the rate of each individual nuclear process within uncertainties to explore the role played by a particular process, or, simultaneously varying the rates of multiple processes (using a Monte Carlo approach) to examine the possible effects of correlated uncertainties. These methods have been used successfully in the past to examine nucleosynthesis in, for example, classical nova explosions (Iliadis et al., 2002; Hix et al., 2003), type II supernovae (The et al., 1998; Jordan, Gupta and Meyer, 2003) and AGB stars (Izzard et al., 2007), and have helped to concentrate experimental efforts on critical reactions identified. For XRBs, progress has been made on sensitivity studies by Amthor et al. (2006) (using the individual-variation approach) and Roberts et al. (2006) (using a Monte Carlo approach). Note that hydrodynamic simulations are computationally prohibitive for large-scale sensitivity studies; hence, one-zone post-processing calculations are generally used.

To address the need for reliable predictions of the composition of the neutron star crust and envelope, we have performed a comprehensive series of investigations into the sensitivity of XRB nucleosynthesis to uncertainties in the input nuclear physics. Both techniques mentioned above (individual-variation and Monte Carlo) were adopted in our studies to allow for comparison of the results, especially in light of the highly-coupled environment expected to develop in these explosive environments (e.g. Roberts et al. (2006)). Our framework was that of post-processing: we used a large, detailed network (606 isotopes up to $^{113}$Xe, linked by 3551 nuclear processes – including weak interactions) in conjunction with a set of thermodynamic histories and initial conditions (Models) to sample the parameter space in XRB nucleosynthesis calculations.

The nuclear network was assembled by using experimental reaction rates when available (e.g. Angulo et al. (1999); Iliadis et al. (2001)), along with theoretical Hauser-Feshbach rates (Rauscher and Thielemann, 2000; Arnould and Goriely, 2006). All reaction rates incorporate the effects of stellar excitations in the target nuclei (Rauscher and Thielemann, 2000). Reactions likely to develop an equilibrium between the forward and reverse processes (namely, reactions with small Q-values – these characterize 'waiting points' in the nuclear reaction activity) were entered using rates calculated with extrapolated proton separation energies from the evaluation of Audi et al. (2003a), when experimental information was unavailable. Mass models may also be used to determine these critical proton separation energies about 'waiting points' – the effects of using different mass models has been explored in Schatz et al. (1998). As well, we have used laboratory beta-decay rates from Audi et al. (2003b).

The Models used in our post-processing calculations arose both from the literature and through parameterization. We used thermodynamic histories (and initial compositions) based upon those of Koike et al. (2004) (K04), Schatz et al. (2001) (S01), and Fisker et al. (2008) (F08). To probe the effects of burst duration and peak temperature achieved, we also scaled the model of Koike et al. (2004) in time (short, long) and temperature (lowT, hiT), respectively (while preserving other parameters in each case). Finally, to evaluate the impact of initial composition, we used the profiles of Koike et al. (2004) with additional, different initial metallicities (lowZ, hiZ). Together, these Models examine XRB conditions as diverse as $T_{peak}$ = 0.9 – 2.5 GK, burst duration ~ 10 -1000 s, and Z = $10^{-4}$ – 0.19.

## Results

We begin by discussing results from individually varying the rate of each nuclear process in our network by an overall factor of 10 up and down, for each Model (~ 40 000 nuclear network calculations). More details on all our results can be found in Parikh et al. (2008). When rates were varied in this manner, a very limited number of reactions were found to significantly affect XRB yields (here, by at least a factor of two relative to a calculation using standard rates) of a large number of isotopes (here, more than three isotopes), in at least one of the Models examined. These reactions are summarized in Table 1. No reaction in Table 1 has sufficient information available for an experimental rate determination over XRB temperatures. Comparing different theoretical estimates of these rates (calculations using different Hauser-Feshbach codes, different Q-values, shell-model calculations – see e.g. Rauscher and Thielemann, (2000); REACLIBv0 database[2]), we find differences as large as a factor ~10 (but usually not more than a factor of 2). The magnitude of these discrepancies helps to support our choice to vary rates by a factor of 10, as opposed to a significantly larger factor (e.g. Amthor et al. (2006)). We note that the 3$\alpha$, $^{12}C(\alpha,\gamma)^{16}O$ and $^{18}Ne(\alpha,p)^{21}Na$ rates were also found to be important according to the above criteria; however, varying these (experimentally known) rates by more realistic uncertainties of 40% for the 3$\alpha$ and a factor of 3 for the other two reactions resulted in only minimal effects on yields. As well, when beta-decay rates were individually varied by their uncertainties as given in Audi et al. (2003b) (usually < 30%), no decay rate was found to significantly affect yields of any isotopes, in any Model.

To examine the impact on XRB yields from the interplay of many reaction rate uncertainties, we used a Monte Carlo approach where all rates were simultaneously varied for 10000 iterations, for each Model. For each iteration, multiplicative factors for the rates were drawn from log-normal distributions; the means of these distributions were set to unity, and we fixed the probability for these factors to fall between 0.1 and 10 as 95.5%. These choices allowed us to make at least a rough comparison with the results of the individual-variation study (in which rates were varied by a fixed factor of 10 up and down). To facilitate such a comparison, we searched for correlations between the final abundance of an isotope X and the multiplicative factor applied to a reaction Y for all X and Y in our network. (Note that multiplicative factors were neither applied to beta-decay rates nor to the 3α reaction as they are known to significantly better precision than a factor of ~10. Tests showed that varying these very important processes by such artificially large factors in the Monte Carlo routine overwhelmed the contribution of uncertainties in other rates to the uncertainties in the final XRB yields.) For each isotope, we then identified those rates that changed the final XRB abundance of that isotope (relative to that from a calculation with standard rates) by at least a factor of two when multiplicative factors between 0.1 and 10 were applied to those rates. In general, the rates found to be important through such an analysis of our Monte Carlo results agreed very well with rates found to be significant in the individual-variation study, for each isotope and Model. Discrepancies were seen for $A \gtrsim 90$; namely, some reactions were indeed seen to affect the final abundances of these isotopes by at least a factor of two in the individual-variation studies, whereas often no such reactions were identified from the analysis of the Monte-Carlo results. This is not a serious issue as for the majority of Models (except S01 and long) the principal nucleosynthesis products are found below A ~ 80. Nonetheless, these discrepancies may demonstrate the effects of correlated uncertainties, which cannot be examined by individual-variation sensitivity studies.

---

[2] http://www.nscl.msu.edu/~nero/db/

Finally, we have explored the effects of uncertainties in reaction Q-values on XRB yields. Reactions with small ($\lesssim$ 1 MeV) Q-values are of particular interest as they quickly achieve equilibrium between the forward and reverse processes in XRB conditions, and therefore represent waiting points during a burst for a continuous reaction flow toward heavier-mass nuclei. To determine the most influential Q-value uncertainties, then, we have examined the effects of individually varying all reactions with $Q < 1$ MeV in our network, by their respective Q-value uncertainties as given in Audi et al. (2003a), for all Models. This involved re-calculating reverse rates for these reactions using $Q+\Delta Q$ and $Q-\Delta Q$. A total of 111 reactions were varied in this manner. Note that experimental information for most of these reaction Q-values is not available; for these cases, we have adopted the extrapolated values (and uncertainties) of Audi et al. (2003a). We find that the uncertainty in the Q-value of the $^{64}$Ge(p,γ)$^{65}$As reaction is overwhelmingly the most critical in determining XRB yields, affecting the yields of isotopes from $^{64}$Zn – $^{104}$Ag by a factor of 2 or more, in most of the Models. The mass of $^{64}$Ge has been measured (Clark et al., 2007; Schury et al., 2007), but the mass of $^{65}$As has only been estimated (as $\Delta$ = -46981 (302) keV in Audi et al. (2003a), where $\Delta$ is the mass excess). The principal waiting-points in XRBs are thought to be $^{64}$Ge, $^{68}$Se and $^{72}$Kr (e.g. Schatz et al. (1998)), and our studies indicate that the Q-value of the $^{64}$Ge(p,γ)$^{65}$As reaction is perhaps the most significant.

Experimental determinations of the rates in Table 1 (or of parameters to improve theoretical estimates of these rates, such as Q-values), as well as the mass of $^{65}$As, would help to better constrain our studies of XRB nucleosynthesis.

## Acknowledgements

This work has been supported by the Spanish MEC grant AYA2007-66256, by the E.U. FEDER funds, and by the U.S. Department of Energy under Contract No. DE-FG02-97ER41041.

**Table 1:** Important reactions found by individually varying each nuclear process in our network by a factor of 10, up and down. Insufficient, if any, experimental information is available for these rates over XRB temperatures. All Q-values are from Audi et al. (2003a). See text for details and Model abbreviations.

| Reaction | Q-value (keV) | Models affected |
|---|---|---|
| $^{25}Si(\alpha,p)^{28}P$ | 6119(11) | hiZ |
| $^{26g}Al(\alpha,p)^{29}Si$ | 4820.68(6) | F08 |
| $^{29}S(\alpha,p)^{32}Cl$ | 5306(50) | hiZ |
| $^{30}P(\alpha,p)^{33}S$ | 1521.36(34) | hiZ |
| $^{30}S(\alpha,p)^{33}Cl$ | 2077(3) | hiZ |
| $^{31}Cl(p,\gamma)^{32}Ar$ | 2422(50) | short |
| $^{32}S(\alpha,\gamma)^{36}Ar$ | 6640.76(14) | long |
| $^{56}Ni(\alpha,p)^{59}Cu$ | -2411(11) | S01, hiZ |
| $^{57}Cu(p,\gamma)^{58}Zn$ | 2277(52) | F08 |
| $^{59}Cu(p,\gamma)^{60}Zn$ | 5120(11) | S01, hiZ |
| $^{61}Ga(p,\gamma)^{62}Ge$ | 2442(149)[a] | F08, short, long, hiZ, lowT |
| $^{65}As(p,\gamma)^{66}Se$ | 2030(424)[a] | K04, short, long, lowZ, hiZ, lowT |
| $^{69}Br(p,\gamma)^{70}Kr$ | 2489(399)[a] | hiT |
| $^{75}Rb(p,\gamma)^{76}Sr$ | 4311(38) | long |
| $^{82}Zr(p,\gamma)^{83}Nb$ | 2055(387)[a] | lowT |
| $^{84}Zr(p,\gamma)^{85}Nb$ | 2946(297)[a] | long |
| $^{84}Nb(p,\gamma)^{85}Mo$ | 4513(409)[a] | lowT |
| $^{85}Mo(p,\gamma)^{86}Tc$ | 1393(409)[a] | F08 |
| $^{86}Mo(p,\gamma)^{87}Tc$ | 1855(530)[a] | F08, lowT |
| $^{87}Mo(p,\gamma)^{88}Tc$ | 2304(300)[a] | lowT |
| $^{92}Ru(p,\gamma)^{93}Rh$ | 2054(499)[a] | long, lowT |
| $^{93}Rh(p,\gamma)^{94}Pd$ | 4467(566)[a] | long |
| $^{96}Ag(p,\gamma)^{97}Cd$ | 3321(566)[a] | K04, long, lowZ, hiT |
| $^{102}In(p,\gamma)^{103}Sn$ | 3554(318)[a] | K04, lowZ |
| $^{103}In(p,\gamma)^{104}Sn$ | 4281(107) | lowZ, hiT |
| $^{103}Sn(\alpha,p)^{106}Sb$ | -5508(432)[a] | S01 |

[a]Q-value and uncertainty estimated from systematic trends. See Audi et al. (2003a).